\documentclass[prd,aps,twocolumn,showpacs,amsmath,amssymb,superscriptaddress,floatfix,nofootinbib,9pt]{revtex4}
\usepackage[colorlinks=true, citecolor=blue, linkcolor=blue, urlcolor=blue]{hyperref}
\usepackage{xcolor,multirow,amsfonts,graphicx,amsfonts,multirow,amsmath}
\usepackage{ulem}

\newcommand{\DS}[1]{/\!\!\!#1}
\newcommand\grap[2]{\includegraphics[width=#1\textwidth]{#2}}

\allowdisplaybreaks
\date{\today}

\begin{document}
\title{Further study on the production of $P$-wave doubly heavy baryons from $Z$-boson decays}
\author{Hai-Jiang Tian$^*$}
\address{Department of Physics, Guizhou Minzu University, Guiyang 550025, P. R. China}
\author{Xuan Luo\footnote{Hai-Jiang Tian and Xuan Luo contributed equally to this work.}}
\address{Department of Physics, Southeast University, Nanjing 210094, P. R. China}
\author{Hai-Bing Fu}
\email{fuhb@alu.cqu.edu.cn (corresponding author)}
\address{Department of Physics, Guizhou Minzu University, Guiyang 550025, P. R. China}

\begin{abstract}
In this paper, we carried out a systematic investigation for the excited doubly heavy baryons production in $Z$-boson decays within the NRQCD factorization approach. Our investigation accounts for all the $P$-wave intermediate diquark states, {\it i.e.} $\langle cc\rangle[^1P_1]_{\bar 3}$, $\langle cc\rangle[^3P_J]_{6}$, $\langle bc\rangle[^1P_1]_{\bar 3/6}$, $\langle bc\rangle[^3P_J]_{\bar 3/6}$, $\langle bb\rangle[^1P_1]_{\bar 3}$, and $\langle bb\rangle[^3P_J]_{6}$ with $J = (0, 1, 2)$. The results show that contributions from all diquark states in $P$-wave were $7\%$, $8\%$, and $3\%$ in comparing with $S$-wave for the production of $\Xi_{cc}$, $\Xi_{bc}$ and $\Xi_{bb}$ via $Z$-boson decay, respectively. Based on these results, we predicted about $0.539\times 10^3(10^6)$ events for $\Xi_{cc}$, $1.827\times 10^3(10^6)$ events for $\Xi_{bc}$, and $0.036\times 10^3(10^6)$ events for $\Xi_{bb}$ can be produced annually at the LHC (CEPC). Additionally, we plot the differential decay widths of $\Xi_{cc}$, $\Xi_{bc}$ and $\Xi_{bb}$ as a function of the invariant mass $s_{23}$ and energy function $z$ distributions, and analyze the theoretical uncertainties in decay width arising from the mass parameters of heavy quark.
\end{abstract}

\date{\today}

\maketitle

\section{Introduction}\label{section:1}
The doubly heavy baryons, contain two heavy quarks and one light quark, are predicted by the quark model theoretically~\cite{Gell-Mann:1964ewy, Ebert:1996ec, Gerasyuta:1999pc, Itoh:2000um}. Investigation for the properties of doubly heavy baryons has been an hot topic from last century to the present. Therefore, a thorough study on them is essential for the particle physics.  In comparing with other baryons that contain only one or none heavy quark, doubly heavy baryons involve more energy scales such as the heavy quark mass  and nonperturbative QCD scale $\Lambda_{\rm QCD}$, which lead to more complex dynamics and offer a novel/distinctive platform for investigating strong interactions.

During the past few decades, experimentalists have put a lot of effort in searching for doubly heavy baryons. Early in 2002 and 2005,  SELEX Collaboration searched a signal for doubly charmed baryon state $\Xi_{cc}^+$, which is conflicated with theoretical predictions~\cite{SELEX:2002wqn,SELEX:2004lln}. Meahwhile, it is also not been confirmed by Belle, LHCb and BaBar Collaborations~\cite{Belle:2013htj, LHCb:2013hvt, BaBar:2006bab}. The excitement is in 2017, the doubly heavy baryon $\Xi_{cc}^{++}$ was firstly observed via decay channel $\Xi_{cc}^{++} \to \Lambda _c^+(\to  p K^- \pi^+) K^- \pi^+ \pi^+ $~\cite{LHCb:2017iph}. This discovery was first reported by the LHCb Collaboration and later verified through another channel $\Xi_{cc}^{++} \to \Xi_c^+ (\to p K^- \pi^+)\pi^+$~\cite{LHCb:2018pcs,LHCb:2019qed}, which inspired many researchers to explore its properties~\cite{LHCb:2018zpl,LHCb:2019gqy,LHCb:2019epo}. In 2020, the LHCb Collaboration searched doubly heavy baryon $\Xi^0_{bc}$ via $\Xi^0_{bc}\to D^0 p K^- $, but no evidence was found~\cite{LHCb:2020iko}. There is a similar case for doubly heavy baryons $\Xi^0_{bc}$ and $\Omega^0_{bc}$, which are searched by LHCb Collaboration through the $\Xi^0_{bc}(\Omega^0_{bc})\to\Lambda_c^+\pi^-$ and $\Xi^0_{bc}(\Omega^0_{bc})\to\Xi_c^{+}\pi^-$ channels, but no signal evidence appeared~\cite{LHCb:2021xba}. As of yet, $\Xi_{bb}$ has not been experimentally discovered.

At the same time, theorists have also devoted much spirit to investigating the properties of doubly heavy baryon production. There are numerous studies in the list of literature that focus on the direct production of them~\cite{Brodsky:2017ntu, Kiselev:1994pu, Falk:1993gb, Chang:2006xp, Baranov:1995rc, Bodwin:1994jh, Gunter:2001qy, Kiselev:1995xe, Berezhnoy:2006mz, Braguta:2002qu, Braaten:2003vy, Li:2007vy, Yang:2007ep, Bi:2017nzv, Zhang:2011hi, Jiang:2012jt, Jiang:2013ej, Martynenko:2013eoa, Yang:2014tca,Yang:2014ita, Martynenko:2014ola, Lai:2014iji, Koshkarev:2016rci, Koshkarev:2016acq, Groote:2017szb, Yao:2018zze, Chang:2006eu, Chen:2014hqa, Zheng:2015ixa, Chen:2018koh, Berezhnoy:2018krl, Chen:2019ykv, Wu:2019gta}. Meanwhile, the indirect production via top quark, Higgs boson, $W$-boson, and $Z$-boson decays is theoretically researched in Refs.~\cite{Niu:2018ycb, Zhang:2022jst, Niu:2019xuq, Luo:2022jxq, Luo:2022lcj, Ma:2022ger, Ma:2022cgt}.
In which, the process of doubly heavy baryons via $W$-boson, top quark, and Higgs boson decays are systematically investigated within NRQCD factorization approach by Zheng and Ma ~\cite{Zhang:2022jst, Ma:2022ger, Ma:2022cgt}. They have explicitly distinguishing the $P$-wave diquark state's contributions with $S$-wave ones. Based on the annual production of $W$-boson and Higgs boson events at the LHC and Circular Electron-Positron Collider (CEPC), the contributions from $P$-wave diquark states will take the non-negligible contributions in comparing with $S$-wave diquark states, which can predict a considerable number of doubly heavy baryons events from the excited states. Therefore, one can further illustrate that the $P$-wave diquark states contributions are significant in detailed calculations.

The doubly heavy baryon production in $Z$-boson decay provides a good chance to study its relevant mechanisms.
On one hand, a large number of $Z$-boson events can be produced at the LHC ($\sim 10^9$ per year~\cite{Liao:2015vqa}), and the accumulated $Z$-boson events will greatly be improved due the increased collision energy (luminosity) at the upgrades of the HE(L)-LHC. On the other hand, the $Z$-boson events can be produced up to $10^{12}$-order level per year at the CEPC~\cite{CEPCStudyGroup:2018ghi}.
The large number of Z-boson events that can be generated by the LHC (CEPC) will provides an ideal platform for collecting doubly heavy baryons produced by $Z$-boson decay, even the excited doubly heavy baryons.
Moreover, the research on doubly excited doubly baryons can be considered as a complement to its ground-state, which significantly inspired us to make a detailed research about it.
Previously, we make a detailed discussion about the intermediate diquark states in the $S$-wave contributions,   which are based on the production of doubly heavy baryons by $Z$-boson decay~\cite{Luo:2022jxq, Luo:2022lcj}. In this paper, we will study the indirect production of excited doubly heavy baryons via $Z$-boson decay and present an overall analysis of the $P$-wave contributions in comparison with $S$-wave.

\begin{figure*}[t]
\includegraphics[width=0.98\textwidth]{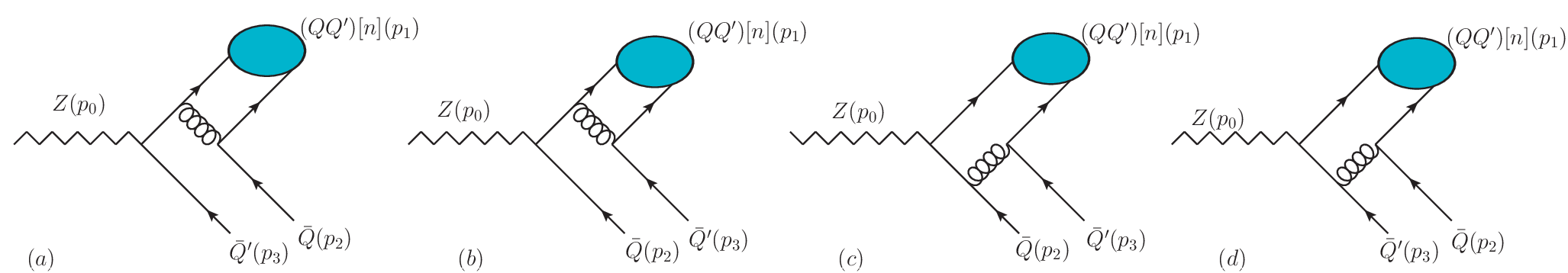}
\caption{The diagrams for $Z \to \langle QQ'\rangle [n] +\bar Q+\bar Q'$ at leading order, where $Q$ and $Q'$ denote the heavy $c$ or $b$-quark respectively.}
\label{fig:1}
\end{figure*}

The remaining parts of the paper are organized as follows: The detailed calculation technology is demonstrated in Sec.~\ref{section:2}. Section~\ref{section:3} presents the phenomenological results and analyses. Finally, a brief summary is provided in Sec.~\ref{section:4}.

\section{Calculation Technology}\label{section:2}
According to the widely accepted hadron production mechanism for doubly heavy baryons $\Xi_{QQ'}$, one can utilize the following two steps to elucidate $\Xi_{QQ'}$ production~\cite{Chang:2006xp, Ma:2003zk, Chang:2006eu, Wu:2019gta}. The first step: When utilizing the decomposition of $SU_c (3)$ color group $3\otimes 3= \bar 3\oplus 6$, the diquark state with possible color quantum number color-antitriplet or color-sextet can be perturbative produced.
The second step: The diquark fragments and then produces an observed double heavy baryon $\Xi_{QQ'q}$ by capturing a light quark from surrounding environment. The fragmentation probability will reach one hundred percent.
In convenience, we take the lable $\Xi_{QQ'}$ to stand for $\Xi_{QQ'q}$ and employ the universal symbol $\langle QQ'\rangle[n]$ to represent a diquark state with color and spin combinations $[n]$ in this paper.
The total ``100\%"  fragmentation probability can further be broken into $43\%, 43\%, 14\%$ probability for $\Xi_{QQ'u}$, $\Xi_{QQ'd}$, $\Xi_{QQ's}$ respectively~\cite{Sun:2020mvl, Chen:2018koh}.

Typical Feynman diagrams for  the process $Z(p_0) \to \langle QQ'\rangle[n](p_1) +\bar Q(p_2)+\bar Q'(p_3)$ at tree level are presented in Fig.~\ref{fig:1}, where the $Q^{(\prime)} $ stand for heavy $c, b$-quark corresponding to the doubly heavy baryons $\Xi_{cc}$, $\Xi_{bc}$ and $\Xi_{bb}$ production. To calculate the differential decay width for $Z (p_0) \to \langle QQ'\rangle[n]({p_1}) +\bar Q(p_2)+\bar Q'(p_3)$ process, one can employ the NRQCD factorization approach~\cite{Bodwin:1996tg,Petrelli:1997ge} and to start with the following formula,
\begin{align}\label{eq:1}
d\Gamma=&\sum_{n}d\hat\Gamma(Z \to \langle QQ'\rangle[n] +\bar Q+\bar Q')\langle {{\mathcal O^H}(n)}\rangle.
\end{align}
In which, long-distance matrix element $\langle {{\mathcal O^H}(n)}\rangle$ indicates the intermediate diquark state $\langle QQ'\rangle[n]$ changed into observable state $\Xi_{QQ'}$. Genarally, it  can be derived from the original value of the Schr\"{o}dinger wave function or its derivative approximately.
Here in this paper, we take the relations $\langle {{\mathcal O^H}(n)}\rangle = (|\Psi_{QQ'}(0)|, |\Psi^{\prime}_{QQ'}(0)|)$ for $S$-wave and $P$-wave doubly heavy baryons, which are mainly derived from experimental data and different non-perturbative theoretical approaches, such as QCD sum rules, lattice QCD, and potential energy models~\cite{Bagan:1994dy, Kiselev:1999sc, Bodwin:1996tg}.

Furthermore, the expression $d\hat\Gamma(Z \to \langle QQ'\rangle[n] +\bar Q+\bar Q')$ can be formulated as:
\begin{align}\label{eq:2}
&d\hat\Gamma(Z \to \langle QQ'\rangle[n] +\bar Q+\bar Q')=\frac13 \frac1{2m_z}{\sum |M[n]|^2}d\Phi_3,
\end{align}
where $ m_Z $ refers to $Z$-boson mass, $|M[n]|$ represent the hard amplitude expressions of the process for the $Z$-boson decay into $S$-wave and $P$-wave doubly heavy baryons, the constant $1/3$ is coming from spin average of initial $Z$-boson, and symbol ``$\sum$'' stands for sum of all final particles' color and spin. $d\Phi_3$ is three-body phase space, which is given by
\begin{eqnarray}
d\Phi_3 = (2\pi)^4 \delta^4 \bigg(p_0 - \sum\limits_f^3 p_f\bigg) \prod \limits_f^3 \frac{d^3p_f}{(2\pi)^3 2p_f^0}~.
\label{eq:3}
\end{eqnarray}
A detailed discussion about the three-body phase space calculations can be found in Refs.~\cite{Chang:2007si,Wu:2008cn}. Subsequently, one can revise the Eq.(\ref{eq:2}) expression as follows:
\begin{eqnarray}
&&d\hat\Gamma(Z\to\langle QQ'\rangle[n] + \bar Q + \bar Q')
\nonumber\\
&&\hspace{2cm} = \frac1{2^8\pi^3 m_z^3} \sum |M[n]|^2 ds_{12} ds_{23}.
\label{eq:4}
\end{eqnarray}
with the invariant mass $s_{ij} = (p_i + p_j)^2$.

\subsection{ Amplitudes for the diquark production}
By applying the charge parity $C=-i\gamma^2\gamma^0$, one can derive the hard amplitude expressions $M[n]$ forbaryon production, which can also be obtained from the process of the $Z$-boson decay into a meson ($Q\bar Q'$)~\cite{Jiang:2012jt,Zheng:2015ixa}.
Here, we have a notation that the charge parity $C$ can be used to reverse one fermion line, which is represented by $L_1=\bar u_{s_1}(k_{12}){\Gamma_{i+1}} S_F(q_i,m_i) \cdots S_F(q_1,m_1)\Gamma_1 v_{s_2}(k_2)$. Where $\Gamma_i$ with index $i=(0,1,...)$ stands for the interaction vertex, $S_F (q_i,m_i)$ is fermion propagator, $s_{1,2}$ is spin index. Then one can obtain:
\begin{align}
&v_{s_2}^{\rm T}(p)C=-\bar \mu_{s_2}(p),&& C^{-1}\Gamma_i^{\rm T} C=-\Gamma_i,
\nonumber\\
&CC^{-1}=I,&&C^{-1}S_f^{\rm T} (-q_i,m_i)C=S_f(q_i,m_i),
\nonumber\\
&C^{-1}\bar \mu_{s_1}^{\rm T} (p_{12})=\nu_{s_1}(p_{12}), && C^{-1}(\gamma^{u})^{\rm T} C=-\gamma^{u},
\nonumber\\
&C^{-1}(\gamma^{u}\gamma^{5})^{\rm T} C=\gamma^{u}\gamma^{5}.
\label{eq:7}
\end{align}
When the axial vector vertices are not included in the fermion line, we can derive:
\begin{eqnarray}
&&\hspace{-0.5cm}L_1 = L_1^{\rm T} = v_{{s_2}}^T(p_2)\Gamma_1^{\rm T} S_F^{\rm T} (q_1,m_1) \cdots S_F^{\rm T} (q_1,m_1)\Gamma_{i+1}^{\rm T} \bar u _{s_1}^{\rm T}({p_{12}})
\nonumber\\
&&= v_{s_2}^{\rm T} (p_2)C C^{-1}\Gamma_1^{\rm T} C C^{-1} S_F^{\rm T} (q_1,m_1)C C^{-1}
\nonumber\\
&&\quad\cdots C{C^{ - 1}}S_F^{\rm T} (q_1,m_1)C C^{-1}\Gamma_{i+1}^{\rm T} C C^{-1}\overline u_{s_1}^{\rm T} (p_{12})
\nonumber\\
&&= (-1)^{i+1}\bar u_{s_2}(p_2)\Gamma_1 S_F(-q_1,m_1)
\nonumber\\
&&\quad\cdots S_F(-q_i,m_i){\Gamma_{i+1}}{v_{s_1}}({p_{12}}).
\label{eq:5}
\end{eqnarray}
For another case that axial vector vertices include fermion line, amplitudes for baryon production can be derived from the similar meson production. But there should multiply an additional factor $(-1)^{(n+1)}$ or $(-1)^{(n+2)}$ to the pure vector case and the axial vector case, respectively.
Therefore, the $S$-wave or $P$-wave amplitude for $Z\to \langle QQ'\rangle[n]+\bar Q+\bar Q'$ decay process can be expressed as:
\begin{eqnarray}
M_{\rm diquark}=(M^a_1-M^v_1)+(M^a_2-M^v_2)+M_3+M_4.
\end{eqnarray}
$M_i$ with $i =(1,2,3,4)$ represent $S$-wave or $P$-wave amplitude of the similar meson production. Specifically, $M^{a,v}_i$ is the axial vector or pure vector part for $M_i$, respectively.
In order to obtain the $S$-wave amplitudes $M_l[n]$ with $l=(a,b,c,d)$, one can obey the Feynman rules based on diagrams shown in Fig.~\ref{fig:1}, which have the following expressions.
\begin{widetext}
\begin{eqnarray}
&&M_a = - \kappa \frac{\bar u(p_{12})(- i\gamma^\nu) v (p_2)\bar u (p_{11}) (- i\gamma^\nu) (m_{Q} + \DS p_1 + \DS p_2)\DS \epsilon(p_0) (c_v^Q  + c_a^Q \gamma^5) v (p_3)}{(p_{12} + p_2)^2 [(p_1 + p_2)^2 - m_Q^2 ]},
\nonumber\\
&&M_b = - \kappa \frac{\bar u (p_{12}) (- i \gamma^\nu ) (m_b + \DS p_1 + \DS p_3)\DS\epsilon(p_0)(c_v^Q  + c_a^Q \gamma^5) v(p_2)\bar u( p_{11}) (- i \gamma^\nu)v (p_3)}{(p_{11} + p_3)^2 [(p_1+ p_3)^2 - m_b^2]},
\nonumber\\
&&M_c = - \kappa \frac{\bar u (p_{12})\DS \epsilon(p_0) (c_v^Q  + c_a^Q \gamma^5) (m_b - \DS p_{11} - \DS p_2 - \DS p_3 ) (-i\gamma^\nu ) v(p_2) \bar u(p_{11}) (-i\gamma^\nu)v (p_3)} {(p_{11} + p_3)^2 [(p_{11} + p_2 + p_3 )^2 - m_b^2]},
\nonumber\\
&&M_d = - \kappa \frac{\bar u (p_{12}) ( - i \gamma^\nu )v(p_2) \bar u(p_{11}) \DS \epsilon(p_0) (c_v^Q  + c_a^Q \gamma^5)(m_{Q} - \DS p_{12} - \DS p_2 - \DS p_3)(-i\gamma^\nu)v(p_3)} {(p_{12}+ p_2 )^2 [(p_{12} + p_2 + p_3 )^2 - m_Q^2]}.
\label{eq:3x}
\end{eqnarray}
In these expressions, the combined factor $\kappa$ have the relation $\kappa=-C g_s^2$, which $C$ is the abbreviation of color factor $C_{ij,k}$.
The $p_{11,12}$ represents momentum for each individual quark in the diquark state. More explicitly, we take $p_{11} = \frac{m_c}{M_{QQ'}} p_1 + q$ and $p_{12} = \frac{m_c}{M_{QQ'}} p_1 - q$, which $q$ is the small relative momentum between different heavy quark $Q^{(\prime)}$.
To maintain gauge invariance, we adopt $M_{QQ'} \simeq m_Q + m_{Q'}$,  and symbol $c_{v,a}^Q$ is the vector or axial vector of $Z_{Q\bar Q}$ vertex coupling constant. It can be written as
\begin{align}
&c_v^c =- \frac{e(8\sin^2\theta_w-3)}{12\cos\theta_w \sin\theta_w},&&c_a^c =- \frac{e}{4\cos\theta_w \sin\theta_w},
\nonumber\\
&c_v^b = \frac{e(4\sin^2\theta_w-3)}{12\cos\theta_w \sin\theta_w},&&c_a^b = \frac{e}{4\cos\theta_w \sin\theta_w}.
\end{align}
In which, the $\theta_w$ stand for Weinberg angle.
After introducing the spin projector $\Pi_{p_1}^{[n]}$, the $S$-wave amplitude can be further expressed as
\begin{eqnarray}
&M_a &= - \kappa \frac{\bar u(p_2)(-i\gamma^\nu )\Pi_{p_1}  (-i\gamma^\nu )(m_{Q} + \DS p_1 + \DS p_2) \DS \epsilon(p_0) (c_v^Q +c_a^Q \gamma^5) v(p_3)} {(p_{12} + p_2)^2\Big[(p_1 + p_2 )^2 - m_{Q}^2 \Big]},
\nonumber\\
&M_b &= - \kappa \frac{\bar u(p_2) \DS \epsilon(p_0)(c_a^Q \gamma^5-c_v^Q )(m_b - \DS p_1 - \DS p_3)(-i\gamma^\nu)\Pi_{p_1}  (-i\gamma^\nu )v(p_3)} {( p_{11} + p_3)^2((p_1 + p_3 )^2 - m_b^2 )},
\nonumber\\
&M_c &=  - \kappa \frac{\bar u( p_2 )(-i\gamma^\nu )(m_b + \DS p_{11} + \DS p_2 + \DS p_3 )\DS \epsilon(p_0) (c_a^Q \gamma^5-c_v^{Q}) \Pi_{p_1}  (-i\gamma^\nu )v(p_3)} {(p_{11} + p_3)^2 \Big[(p_{11} + p_2 + p_3)^2 - m_b^2 \Big]},
\nonumber\\
&M_d &= - \kappa \frac{{\bar u( {{p_{2}}} )(-i\gamma^\nu )\Pi_{p_1}  \DS \epsilon(p_0) (c_v^Q  + c_a^Q \gamma^5)(m_{Q'} - \DS p_{12} - \DS p_2 - \DS p_3} )(-i\gamma^\nu )v(p_3)} {(p_{12} + p_2 )^2 \Big[(p_{12} + p_2 + p_3 )^2 - m_{Q}^2\Big]}.
\label{eq:3}
\end{eqnarray}
\end{widetext}
The spin projector $\Pi_{p_1} $ appear in Eq.~\eqref{eq:3} can be expressed as~\cite{Bodwin:2002cfe}
\begin{eqnarray}
\Pi_{p_1}(q) = \frac{-\sqrt{M_{QQ'}}}{4{m_{Q}m_{Q'}}} (\DS p_{12}-m_Q)\gamma^5(\DS p_{12} + m_Q),
\label{eq:8}
\end{eqnarray}
or
\begin{eqnarray}
\Pi^{\beta}_{p_1}(q) = \frac{-\sqrt{M_{QQ'}}}{4{m_{Q}m_{Q'}}} (\DS p_{12}-m_Q)\gamma^{\beta}(\DS p_{12} + m_Q),
\label{eq:48}
\end{eqnarray}
which stands for spin-singlet state or spin-triplet state, respectively.

On the other hand, the $P$-wave diquark states amplitudes can be obtained by making the derivative for $S$-wave expression either a spin singlet or a spin triplet.
Here we have a notation that the doubly heavy baryons can be excited into $P$-wave state either in $\rho$-mode or $\lambda$-mode, which is from the excitation between $Q$ and $Q'$ or excitation between $QQ'$ and $q$. In this paper, we solely focus on the effect of $\rho$-mode to make investigation.
\begin{widetext}
\begin{eqnarray}
&M_a[n]&=\epsilon^l_a(q)\frac{d}{dq_a} \Big[- \kappa \frac{\bar u(p_2)(-i\gamma^\nu )\Pi_{p_1}  (-i\gamma^\nu )(m_{Q} + \DS p_1 + \DS p_2) \DS \epsilon(p_0) (c_v^Q +c_a^Q \gamma^5) v(p_3)} {(p_{12} + p_2)^2((p_1 + p_2 )^2 - m_{Q}^2 )}\Big],
\nonumber\\
&M_b[n]&= \epsilon^l_a(q)\frac{d}{dq_a} \Big[- \kappa \frac{\bar u(p_2) \DS \epsilon(p_0)(c_a^Q \gamma^5-c_v^Q )(m_b - \DS p_1 - \DS p_3)(-i\gamma^\nu)\Pi_{p_1}  (-i\gamma^\nu )v(p_3)} {( p_{11} + p_3)^2((p_1 + p_3 )^2 - m_b^2 )}\Big],
\nonumber\\
&M_c[n]&=\epsilon^l_a(q)\frac{d}{dq_a} \Big[  - \kappa \frac{\bar u( p_2 )(-i\gamma^\nu )(m_b + \DS p_{11} + \DS p_2 + \DS p_3 )\DS \epsilon(p_0) (c_a^Q \gamma^5-c_v^{Q}) \Pi_{p_1}  (-i\gamma^\nu )v(p_3)} {(p_{11} + p_3)^2 ((p_{11} + p_2 + p_3)^2 - m_b^2 )}\Big],
\nonumber\\
&M_d[n]&= \epsilon^l_a(q)\frac{d}{dq_a} \Big[- \kappa \frac{{\bar u( {{p_{2}}} )(-i\gamma^\nu )\Pi_{p_1}  \DS \epsilon(p_0) (c_v^Q  + c_a^Q \gamma^5)(m_{Q'} - \DS p_{12} - \DS p_2 - \DS p_3} )(-i\gamma^\nu )v(p_3)} {(p_{12} + p_2 )^2 ((p_{12} + p_2 + p_3 )^2 - m_{Q}^2)}\Big].
\label{eq:3}
\end{eqnarray}
and
\begin{eqnarray}
&M_a[n]&= \epsilon^l_{a\beta}(q)\frac{d}{dq_a} \Big[- \kappa \frac{\bar u(p_2)(-i\gamma^\nu )\Pi^{\beta}_{p_1}  (-i\gamma^\nu )(m_{Q} + \DS p_1 + \DS p_2) \DS \epsilon(p_0) (c_v^Q +c_a^Q \gamma^5) v(p_3)} {(p_{12} + p_2)^2((p_1 + p_2 )^2 - m_{Q}^2 )}\Big],
\nonumber\\
&M_b[n]&=\epsilon^l_{a\beta}(q)\frac{d}{dq_a} \Big[ - \kappa \frac{\bar u(p_2) \DS \epsilon(p_0)(c_a^Q \gamma^5-c_v^Q )(m_b - \DS p_1 - \DS p_3)(-i\gamma^\nu)\Pi^{\beta}_{p_1}  (-i\gamma^\nu )v(p_3)} {( p_{11} + p_3)^2((p_1 + p_3 )^2 - m_b^2 )}\Big],
\nonumber\\
&M_c[n]&=  \epsilon^l_{a\beta}(q)\frac{d}{dq_a} \Big[- \kappa \frac{\bar u( p_2 )(-i\gamma^\nu )(m_b + \DS p_{11} + \DS p_2 + \DS p_3 )\DS \epsilon(p_0) (c_a^Q \gamma^5-c_v^{Q}) \Pi^{\beta}_{p_1}  (-i\gamma^\nu )v(p_3)} {(p_{11} + p_3)^2 ((p_{11} + p_2 + p_3)^2 - m_b^2 )}\Big],
\nonumber\\
&M_d[n]&= \epsilon^l_{a\beta}(q)\frac{d}{dq_a} \Big[ -\kappa \frac{{\bar u( {{p_{2}}} )(-i\gamma^\nu )\Pi^{\beta}_{p_1}  \DS \epsilon(p_0) (c_v^Q  + c_a^Q \gamma^5)(m_{Q'} - \DS p_{12} - \DS p_2 - \DS p_3} )(-i\gamma^\nu )v(p_3)} {(p_{12} + p_2 )^2 ((p_{12} + p_2 + p_3 )^2 - m_{Q}^2)}\Big].
\label{eq:3}
\end{eqnarray}
\end{widetext}
In which, the $\epsilon^l_{\alpha}(q)$ and $\epsilon^l_{\alpha\beta}(q)$ are the polarized vector and polarized tensor for spin singlet and spin triplet of the $P$-wave diquark state.
After sum over the $[^1P_1]$ states polarized vectors, one can get the following result
\begin{align}\label{align1}
\sum\limits_{{l_z}} {\varepsilon_\alpha ^l\varepsilon _{\alpha'}^{l*} = \Pi {_{\alpha \alpha '}}}.
\end{align}
In which the symbol $l_z$ represents the excited states $[^1P_1]$. In the case of excited states $[^3P_J]$, the sum over polarized tensors are represented by
\begin{align}\label{align2}
\varepsilon _{\alpha \beta }^0\varepsilon _{\alpha '\beta '}^{0*} &= \frac{1}{3}\Pi {_{\alpha \alpha '}}, \\ \nonumber
\sum\limits_{{J_z}} {\varepsilon _{\alpha \beta }^1\varepsilon _{\alpha '\beta '}^{1*}}  &= \frac{1}{2}(\Pi {_{\alpha \alpha '}} \Pi {_{\beta \beta '}}  - \Pi {_{\alpha \beta '}} \Pi {_{\beta \alpha '}} ),\\ \nonumber
\sum\limits_{{J_z}} {\varepsilon _{\alpha \beta }^2\varepsilon _{\alpha '\beta '}^{2*}}  &= \frac{1}{2}(\Pi {_{\alpha \alpha '}} \Pi {_{\beta \beta '}}  + \Pi {_{\alpha \beta '}} \Pi {_{\alpha '\beta }} ) \\ \nonumber
&- \frac{1}{3}\Pi {_{\alpha \beta '}} \Pi {_{\beta \alpha '}},
\end{align}
with the definition
\begin{align}\label{align3}
\Pi {_{\alpha \beta } =  - {g_{\alpha \beta }} - \frac{{{p_{1\alpha}}{p_{1\beta }}}}{{M_{QQ'}^2}}}.
\end{align}
\subsection{ Color Factor }
According to Fig.~\ref{fig:1}, it is easy to write the expression for color factor $C_{ij,k}$,
\begin{eqnarray}\label{eq:10}
C_{ij,k} = N \times \sum\limits_{a,m,n} (T^a)_{im}(T^a)_{jn}\times G_{mnk},
\end{eqnarray}
with letters $k$ and $a$ are color indices of diquark and gluon, respectively, and $N = \sqrt {1/2}$ refers to the normalization factor. In addition, those marks $i,j,m$, and $n=(1,2,3)$ are also the color indices, which correspond to two outgoing antiquarks and two constituent active quarks in the diquark state, respectively. For the $\bar 3(6)$ state, the value of $G_{mnk}$ is related to the function  $\varepsilon_{mjk}(f_{mjk})$, by given as
\begin{eqnarray}\label{eq:11}
&&\varepsilon_{mjk}\varepsilon_{m'j'k} = \delta_{mm'}\delta_{jj'} - \delta_{mj'}\delta_{jm'},
\nonumber\\[2ex]
&&f_{mjk}f_{m'j'k} = \delta_{mm'} \delta_{jj'} + \delta_{mj'}\delta_{jm'}.
\end{eqnarray}

\subsection{Hadronization from diquark $(QQ')[n]$ to $\Xi_{QQ'}$}
The process of diquark states $\langle QQ'\rangle[n]$ hadronization into the final states $\Xi_{QQ'}$ is regarded as a non-perturbative, and its impact is reflected by the overall factors $\left\langle \mathcal O^H(n) \right\rangle$. The factors $\left\langle \mathcal O^H(n) \right\rangle$, also known as the transition probabilities, can be represented by $h_{\bar 3}$ and $h_6$ in color $\bar 3$ and color $6$, respectively. Detailed researches and discussions about $h_{\bar3}$ ($h_{6}$) can be found in Refs.~\cite{Zheng:2015ixa,Ma:2003zk,Jiang:2012jt,Niu:2018ycb,Zhang:2022jst}.

In accordance with the velocity scaling rule of the NRQCD factorization approach, one can assume that the transition probabilities $h_{\bar 3}$ and $h_6$ hold an equal horizon~\cite{Chang:2006xp, Chang:2006eu, Ma:2003zk}. Therefore, the wave function evaluated at the origin can be interconnected with them~\cite{Chang:2006eu}.
\begin{eqnarray}
h_6 \sim h_{\bar3} = \langle \mathcal O^H(n)\rangle  = |\Psi(0)|^2,
\end{eqnarray}
for $S$-wave, and for $P$-wave
\begin{eqnarray}
h_6 \sim h_{\bar3} = \langle \mathcal O^H(n)\rangle  = | \Psi^{\prime}(0)|^2.
\end{eqnarray}
The wave function at the origin can be naturally connected to the radial wave function at the origin:
\begin{eqnarray}
&&|\Psi(0)|^2=\dfrac{1}{4\pi}|R(0)|^2,  \nonumber \\
&&|\Psi'(0)|^2=\dfrac{3}{4\pi}|R'(0)|^2.
\end{eqnarray}
Based on the NRQCD factorization approach, $\Xi_{QQ'}$ is a bound state of two heavy quarks with other light dynamical freedom of QCD and can be can be described by a series of Fock states:
\begin{eqnarray}
\left| {{\Xi _{QQ'}}} \right\rangle& = {c_1}(v)\left| {(QQ')q} \right\rangle +{c_2}(v)\left| {(QQ')qg} \right\rangle
\\ \nonumber
 & + {c_3}(v)\left| {(QQ')qgg} \right\rangle  +  \cdots
 \label{qe1}
\end{eqnarray}
Due to the light quark may produce gluons easily, the constituents $c_i (v)$ with $i=(1,2,3...)$ in Eq.~\eqref{qe1} have equivalent significance with each other, {\it e.g.} $c_1(v)\sim c_2(v)\sim c_3(v)$. In the $\langle QQ'\rangle[^3S_1]_{\bar  3}$ state, one of heavy quarks can produce a gluon without altering its spin, and the gluon can divide into a light quark pair $q\bar q$.
Then the heavy diquark state $\langle QQ'\rangle$ can form a final baryon state by capturing a light quark $q$.
In $\langle QQ'\rangle[^1S_0]_6$ state, the heavy quark spin must be altered by the emitted gluon when a baryon is formed by $\left| {(QQ')q} \right\rangle$. The reason lies in a suppression from $h_6$.
In the case of the heavy quark's spin remaining unchanged, one of the heavy quarks produces a gluon, and this gluon can separate into a quark-antiquark $q\bar q$ pair. Additionally, the light quark $q$  has the ability to produce a gluon, which can be used to construct the component with $qg$.
Based on the above analysis, the $h_6$ and $h_{\bar 3}$ hold the same order in $v_c$. We assume $h_6$ and $h_{\bar 3}$ hold the same order in $v_c$ for convenience.

\section{Numerical Results}\label{section:3}

\begin{table}[t]
\center
\caption{Our results for decay widths (in unit: KeV), branching ratios , and events at the LHC (CEPC) of the process $Z$-boson decay into $\Xi_{cc}$. States denote the intermediate diquark.}
\begin{tabular}{|c|c|c|c|c|}
\hline
State    & Decay width & Br($\times10^{-6}$) & LHC events & CEPC event \\ \hline
$[^3S_1]_{\bar3}$    & 12.698      & $5.089$    & $5.089\times 10^{3} $         & $5.089\times 10^{6}$         \\ \hline
$[^1S_0]_{6}$          & 6.177       & $2.476$    & $2.476\times 10^{3}$          & $2.476\times 10^{6}  $        \\ \hline
$[^1P_1]_{\bar3}$    & 0.475       & $0.190$    & $0.190\times 10^{3}$          & $0.190\times 10^{6}$          \\ \hline
$[^3P_0]_{6}$          & 0.342       & $0.137$    & $0.137\times 10^{3}$          &$0.137\times 10^{ 6}$          \\ \hline
$[^3P_1]_{6}$          & 0.379       & $0.152$    & $0.152\times 10^{3}$          & $0.152\times 10^{ 6}$          \\ \hline
$[^3P_2]_{6}$          & 0.149       & $0.060$    & $0.060\times 10^{3}$          & $0.060\times 10^{6}$          \\ \hline
$S$-wave                   & 18.875      & $7.565$    & $7.565\times 10^{3}$          & $7.565\times 10^{ 6} $         \\ \hline
$P$-wave                   & 1.345       & $0.539$    & $0.539\times 10^{3}$          & $0.539\times 10^{ 6}$          \\ \hline
Total                    & 20.220      & $8.104$    & $8.104\times 10^{3}$          & $8.104\times 10^{ 6}$           \\ \hline
\end{tabular}
\label{tab1}
\end{table}
To calculate the numerical results, the choices of the following input parameters are applied
\begin{align}\label{int}
&m_c=1.8~{\rm GeV},~~~m_b=5.1~{\rm GeV},~~~m_Z=91.1876~{\rm GeV},
\nonumber \\
&M_{\Xi_{cc}}=3.6~{\rm GeV},~~M_{\Xi_{bc}}=6.9~{\rm GeV},~~M_{\Xi_{bb}}=10.2~{\rm GeV},
\nonumber \\
&G_F=1.1663787\times 10^{-5}~{\rm GeV}^{-2},\theta_w={\rm arcsin}\sqrt {0.2312},
\nonumber \\
&\Gamma_{Z}=2.4952.
\end{align}
which are consistent with Refs.~\cite{Baranov:1995rc,ParticleDataGroup:2018ovx}.
As a common choice, the values of $|R(0)|$ and $|R'(0)|$ keep the same with Ref.~\cite{Kiselev:2002iy}.
\begin{eqnarray}
&&|\Psi_{cc}(0)|=0.523~{\rm GeV^{\frac{3}{2}}},~|\Psi'_{cc}(0)|=0.102~{\rm GeV^{\frac{5}{2}}},~~~   \nonumber \\
&&|\Psi_{bc}(0)|=0.722~{\rm GeV^{\frac{3}{2}}},~|\Psi'_{bc}(0)|=0.200~{\rm GeV^{\frac{5}{2}}},~~~ \nonumber \\
&&|\Psi_{bb}(0)|=1.345~{\rm GeV^{\frac{3}{2}}},~|\Psi'_{bb}(0)|=0.479~{\rm GeV^{\frac{5}{2}}}.~~~
\label{int1}
\end{eqnarray}
For the production of $\Xi_{cc, bc}$ and $\Xi_{bb}$ baryons, we take the strong coupling constant $\alpha_s(2m_c) = 0.242$ and $\alpha_s(2m_b) = 0.180$ separately. In the following discussions, contributions from $S$-wave diquark states are also present in Tables~~\ref{tab1}, \ref{tab2} and \ref{tab3} thorough comparison and analysis.

\subsection{The $\Xi_{cc}$ production}\label{section:33}
\begin{figure}[t]
\grap{0.43}{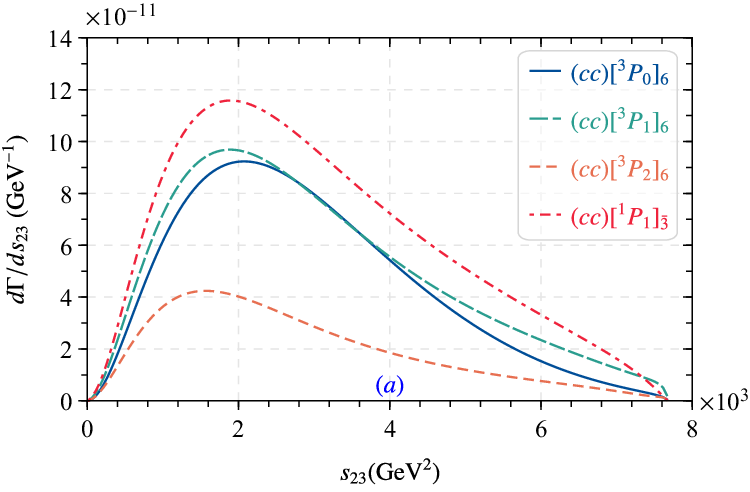}
\grap{0.41}{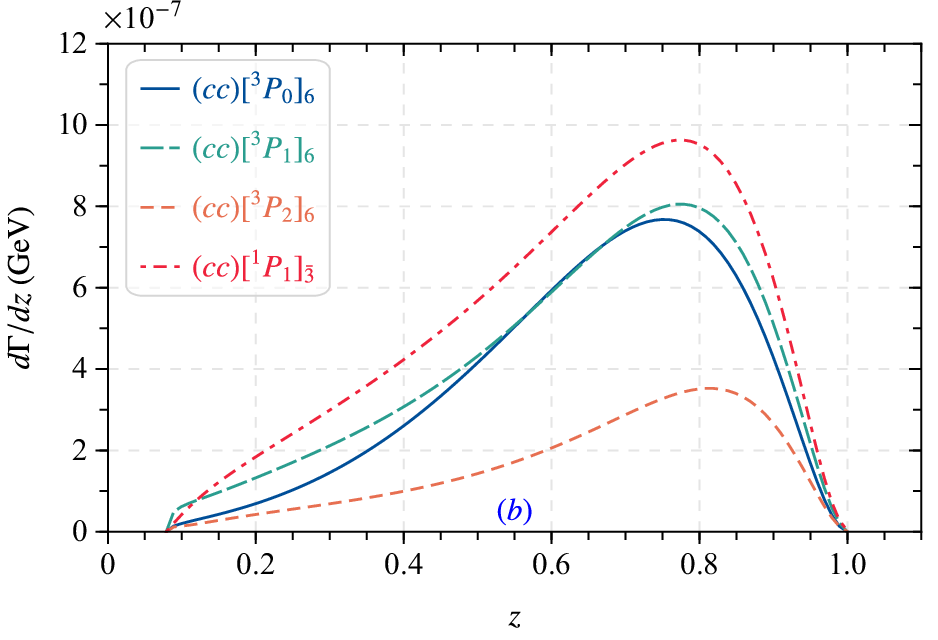}
\caption{The $d\Gamma/ds_{23}$ and $d\Gamma/dz$ for the $Z$-boson decay into $\Xi_{cc} $, where $\bar 3(6)$ is the color quantum number of diquark state.}
\label{fig:ccs}
\end{figure}
At this stage, we present the $Z \to \Xi_{cc}$ decay widths, branching ratios, and events at the LHC (CEPC)  in Table~\ref{tab1}, where the branching ratio has the definition as follows:
\begin{align}\label{ccqe1}
{\rm Br}[n]=\dfrac{\Gamma_{Z\to\Xi_{cc}[n]+\bar c+\bar c}}{\Gamma_Z},
\end{align}
with $[n]$ represents the intermediate spin-color state.Meanwhile, we also put the sum of states $[^1S_0]_6$ and $[^3S_1]_{\bar3}$ labeled as `$S$-wave', excited diquark states $[^1P_1]_{\bar3}$ and $[^3P_J ]_6$ with $J = (0, 1, 2)$ labeled as `$P$-wave' in Table~\ref{tab1}. From the table, one can observe that:
\begin{itemize}
\item  The contribution for $[^1P_1]_{\bar3}$ state is the largest among the excited states in the production of $\Xi_{cc}$  via $Z$-boson decay, and its ratio to the other excited states $[^3P_0]_6$, $[^3P_1]_6$, $[^3P_2]_6$ is $1:0.72:0.79:0.31$.
\item The total contribution from intermediate $P$-wave states is about $7\%$ to the $S$-wave states, which indicate that $P$-wave states play a significant role in detailed calculations.
\item When all the excited states under consideration are combined, the events number from excited states via $Z$-boson decay can reach to a massive amount  in the order $10^3~(10^6)$ at the LHC (CEPC) for one year.
\item At LHC (CEPC), about $1.827\times10^3(10^6)$ $\Xi_{bc}$ events will be produced one year when all the considered excite states are summed up. At the same time, the events for $\langle cc\rangle[^1P_1]_{\bar 3}$, $\langle cc\rangle[^3P_0]_6$, $\langle cc\rangle[^3P_1]_6$, and $\langle cc\rangle[^3P_2]_6$ diquark states are about $0.190\times10^3( 10^6)$, $0.137\times10^3( 10^6)$, $0,152\times10^3( 10^6)$, and $0.060\times10^3( 10^6)$ one year.
\end{itemize}
\begin{table}[t]
\centering
\caption{Our results for decay widths (in unit: KeV), branching ratios , and events at the LHC (CEPC) of the process $Z$-boson decay into $\Xi_{bc}$. States denote the intermediate diquark.}
\begin{tabular}{|c|c|c|c|c|}
\hline
State   & Decay width & Br($\times 10^{-6}$) & LHC events & CEPC event \\ \hline
$[^3S_1]_{\bar3}$    & 21.951      & $8.797$      & $8.797\times 10^{3}$          & $8.797\times 10^{6}$     \\ \hline
$[^3S_1]_{6}$          & 10.976      & $4.399$      & $4.399\times 10^{3}$          & $4.399\times 10^{6}$     \\ \hline
$[^1S_0]_{\bar3}$    & 16.218      & $6.500$      & $6.500\times 10^{3}$          & $6.500\times 10^{6}$      \\ \hline
$[^1S_0]_{6}$          & 8.109       & $3.250$      & $3.250\times 10^{3}$          & $3.250\times 10^{6}$   \\ \hline
$[^1P_1]_{\bar3}$    & 0.720       & $0.289$      & $0.289\times 10^{3}$          & $0.289\times 10^{6}$   \\ \hline
$[^1P_1]_{6}$          & 0.360       & $0.144$      & $0.144\times 10^{3}$          & $0.144\times 10^{6}$    \\ \hline
$[^3P_0]_{\bar3}$    & 0.498       & $0.200$      & $0.200\times 10^{3}$          & $0.200\times 10^{6}$   \\ \hline
$[^3P_0]_{6}$          & 0.249       & $0.100$      & $0.100\times 10^{3}$          & $0.100\times 10^{6}$   \\ \hline
$[^3P_1]_{\bar3}$    & 0.928       & $0.372$      & $0.372\times 10^{3}$          & $0.372\times 10^{6}$    \\ \hline
$[^3P_1]_{6}$          & 0.464       & $0.186$      & $0.186\times 10^{3}$          & $0.186\times 10^{6}$   \\ \hline
$[^3P_2]_{\bar3}$    & 0.893       & $0.358$      & $0.358\times 10^{3}$          & $0.358\times 10^{6}$    \\ \hline
$[^3P_2]_{6}$          & 0.447       & $0.179$      & $0.179\times 10^{3}$          & $0.179\times 10^{6}$   \\ \hline
$S$-waves                   & 57.254      & $22.946$     & $22.946\times 10^{3}$         & $22.946\times 10^{6}$   \\ \hline
$P$-waves                   & 4.559       & $1.827$      & $1.827\times 10^{3}$          & $1.827\times 10^{6}$    \\ \hline
Total                    & 61.813      &$ 24.773$     & $24.773\times 10^{3}$         & $24.773\times 10^{6}$     \\ \hline
\end{tabular}
\label{tab2}
\end{table}
In order to take a deep looking at these channels for $\Xi_{cc}$ production, we have plotted the differential decay width curves with respect to $s_{23}$ and $z$ in Fig.~\ref{fig:ccs}. The $s_{23} = (p_2 + p_3)^2$ stand for the invariant mass. The $z = 2E_1/E_Z$ is the energy fraction, where $E_{1,Z}$ is the energy of $\Xi_{cc}$ or $Z$-boson.
\begin{itemize}
  \item Fig.~\ref{fig:ccs}(a) provides a clear demonstration that the distributions of various excited states are similar with $\Xi_{cc}$ production. These curves increase with $s_{23}$ initially, and then decreased. Their peaks are lie in a small $s_{23}$ region.
  \item Fig.~\ref{fig:ccs}(b) illustrates that the $d\Gamma$ behavior changed with the energy fraction $z$ distribution, {\it i.e.}, $d\Gamma/dz$ are similar with $d\Gamma/ds_{23}$, which are initial increased and then decreased. Their peaks are located at large $z$ region.
\end{itemize}
\subsection{The $\Xi_{bc}$ production}\label{section:33}
Secondly, The decay widths, branching ratios, and events at the LHC (CEPC)  for $Z \to \Xi_{bc}$ are displayed in Table~\ref{tab2},  which has following notations:

\begin{itemize}
\item In excited baryon $\Xi_{bc}$ production via $Z$-boson decays, the largest contribution among the excited states comes from the spin and color state $[^3P_1]_{\bar3}$, with a ratio of $[^1P_1]_{\bar 3}: [^1P_1]_{6}: [^3P_0]_{\bar 3}: [^3P_1]_{\bar 3}:[^3P_2]_{\bar 3}:[^3P_0]_6:[^3P_1]_6 :[^3P_2]_6=1: 0.5: 0.69: 0.35:1.3:0.65:0.62:0.62$.

\item The total intermediate $P$-wave states contribution for $Z$-boson decay into excited baryon $\Xi_{bc}$ is about $7\%$ in comparing with $S$-wave state.

\item  When all the considered excite states are summed up, there will be  $1.827\times10^3(10^6)$ $\Xi_{bc}$ events produced at the LHC (CEPC) for one year. Meanwhile, the event for the diquark states $\langle bc\rangle[^1P_1]_{\bar 3}$, $\langle bc\rangle[^1P_1]_{6}$, $\langle bc\rangle[^3P_0]_{\bar 3}$, $\langle bc\rangle[^3P_1]_{\bar 3}$, $\langle bc\rangle[^3P_2]_{\bar 3}$, $\langle bc\rangle[^3P_0]_6$, $\langle bc\rangle[^3P_1]_6$, and $\langle bc\rangle[^3P_2]_6$ are about $0.289\times10^3( 10^6)$, $0.144\times10^3( 10^6)$, $0,200\times10^3( 10^6)$, $0.100\times10^3( 10^6)$, $0.372\times10^3( 10^6)$, $0.186\times10^3(10^6)$,
     $0.358\times10^3( 10^6)$, $0.179\times10^3( 10^6)$, separately.
\end{itemize}

\begin{figure}[t]
\grap{0.43}{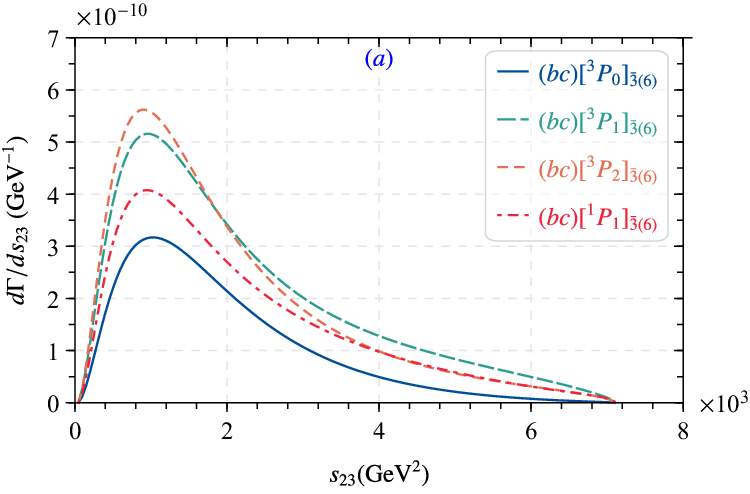}
\grap{0.41}{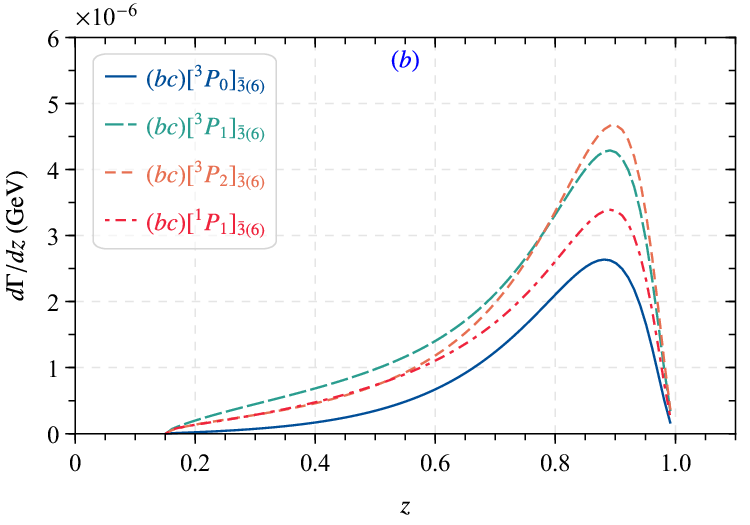}
\caption{The $d\Gamma/ds_{23}$ and $d\Gamma/dz$ for the $Z$-boson decay into $\Xi_{bc} $, where ${\bar 3(6)}$ is the color quantum number of diquark state, and $[n]_{\bar 3(6)}$ means the sum of the results from $[n]_{3}$ and $[n]_6$.}
\label{fig:bcs}
\end{figure}
In order to show the behaviors for process $Z$-boson decay into excited doubly heavy baryon $\Xi_{bc}$, we plotted the differential decay widths with respect to $s_{23}$ and $z$ in Fig.~\ref{fig:bcs}. There are eight states totally for $\Xi_{bc}$ baryon, as both color-antitriplet excited state and color-sextet excited state are reasonable for any spin states. To make these figures more clearer, we sum several colored states with same spin state. For example, the line labeled with $\langle bc\rangle[^1P_1]_{\bar3(6)}$ means the sum of the contributions from $\langle bc\rangle[^1P_1]_{\bar3}$ and $\langle bc\rangle[^1P_1]_{6}$.  The similarities between the angular and invariant mass differential widths of $\Xi_{cc}$ and $\Xi_{bc}$ productions via $Z$-boson decays indicate their similar kinematic behaviors.

\subsection{The $\Xi_{bb}$ production}\label{section:33}
\begin{table}[t]
\centering
\caption{Our results for decay widths (in unit: KeV), branching ratios, and events at the LHC (CEPC) of the process $Z$-boson decay into $\Xi_{bb}$. In which, the states denote the intermediate diquark.}
\begin{tabular}{|c|c|c|c|c|}
\hline
State    & Decay width & Br($\times 10^{-6}$) & LHC events & CEPC event \\ \hline
$[^3S_1]_{\bar3}$    & 1.893     & $0.759$               & $0.759\times 10^{3}$          & $0.759\times 10^{ 6} $   \\ \hline
$[^1S_0]_{6}$          & 0.974     & $0.390$               & $0.390\times 10^{3}$          & $0.390\times 10^{ 6} $  \\ \hline
$[^1P_1]_{\bar3}$    & 0.029     & $0.012$               & $0.012\times 10^{3}$          & $0.012\times 10^{ 6} $ \\ \hline
$[^3P_0]_{6}$          & 0.024     & $0.010$               & $0.010\times 10^{3}$          & $0.010\times 10^{ 6} $ \\ \hline
$[^3P_1]_{6}$          & 0.027     & $0.011$               & $0.011\times 10^{3}$          & $0.011\times 10^{ 6} $\\ \hline
$[^3P_2]_{6}$          & 0.011     & $0.004$               & $0.004\times 10^{3}$          & $0.004\times 10^{ 6} $  \\ \hline
$S$-waves                   & 2.867     & $1.149$               & $1.149\times 10^{3}$          & $1.149\times 10^{ 6} $ \\ \hline
$P$-waves                   & 0.091     & $0.036$               & $0.036\times 10^{3}$          & $0.036\times 10^{ 6} $  \\ \hline
Total                    & 2.958     & $1.185$               & $1.185\times 10^{3}$          & $1.185\times 10^{ 6} $  \\ \hline
\end{tabular}
\label{tab3}
\end{table}

\begin{figure}[t]
\grap{0.43}{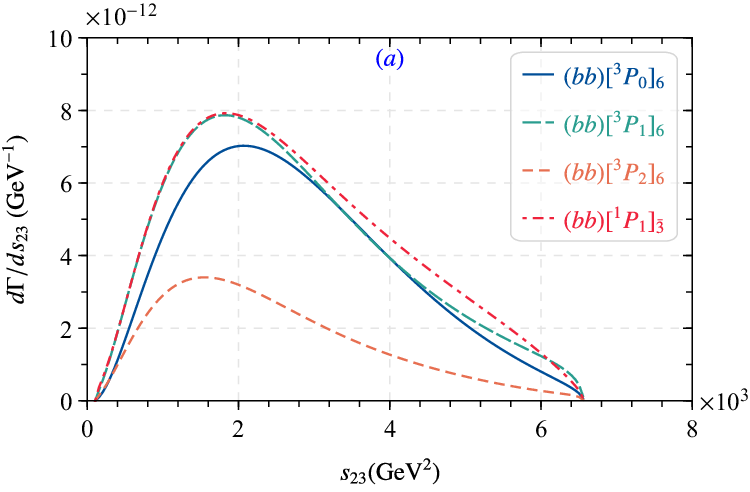}
\grap{0.41}{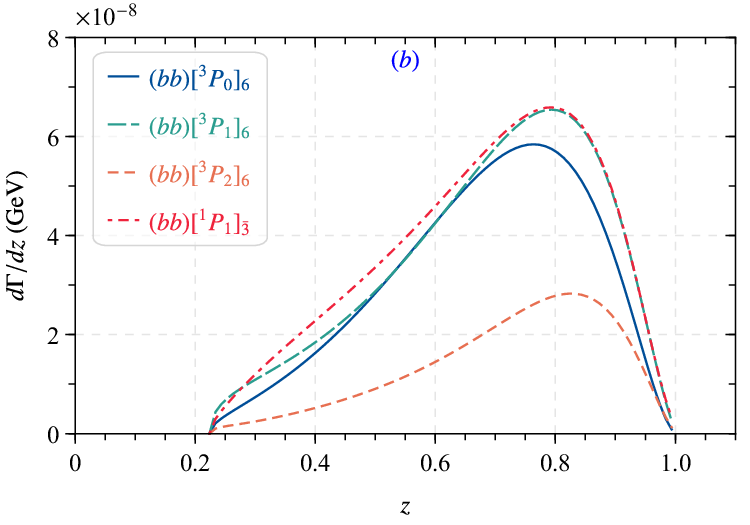}
\caption{The $d\Gamma/ds_{23}$ and $d\Gamma/dz$ for the $Z$-boson decay into $\Xi_{bb} $, where ${\bar 3(6)}$ is the color quantum number of diquark state.}
\label{fig:bbs}
\end{figure}

Thirdly, decay widths, branching ratios, and events at LHC (CEPC) of the $Z$-boson decay into $\Xi_{bb}$ are displayed in Table~\ref{tab3}, which is similar with $\Xi_{cc}$ and $\Xi_{ bc}$ cases. From which we can get the conclusions
\begin{itemize}
\item In the case of the excited state $\Xi_{bb}$ production via $Z$-boson decay, the largest contribution among excited states comes from spin and color state $[^1P_1]_{\bar3}$. The ratio of $\langle bb\rangle[^1P_1]_{\bar 3}$: $\langle bb\rangle[^3P_0]_6$: $\langle bb\rangle[^3P_1]_6$: $\langle bb\rangle[^3P_2]_6=1:0.83:0.93:0.41$.
\item The total contribution from intermediate $P$-wave states is roughly $3\%$ of that from $S$-wave.
\item When all the considered excite states are summed up, about $0.036\times10^3(10^6)$ $\Xi_{bc}$ events per year will be produced at the LHC (CEPC). The events in
    $\langle cc\rangle[^1P_1]_{\bar 3}$, $\langle cc\rangle[^3P_0]_6$, $\langle cc\rangle[^3P_1]_6$, and $\langle cc\rangle[^3P_2]_6$ diquark states are about $0.012\times10^3( 10^6)$, $0.010\times10^3( 10^6)$, $0,011\times10^3( 10^6)$, and $0.004\times10^3( 10^6)$ for one year.
\end{itemize}
The differential decay widths for $\Xi_{bb}$ with respect to $s_{ij}$ and $z$ are plotted in Fig.~\ref{fig:bbs}. From which it can be observed that the characteristics are fundamentally consistent with the production of $\Xi_{cc}$ or $\Xi_{bc}$.

\begin{table}[b]
\centering
\caption{ Our result for decay widths (in unit: KeV) of the process $Z$-boson decay into $\Xi_{cc}$ by varying $m_c$ (in unit: GeV).}
\begin{tabular}{|c|c|c|c|}
\hline
State & $m_c$=1.5GeV & $m_c$=1.8GeV & $m_c$=2.1GeV \\ \hline
 $[^1P_1]_{\bar3}$ & 1.384     & 0.475     & 0.193     \\ \hline
$[^3P_0]_{6}$        & 0.986     & 0.342     & 0.141     \\ \hline
$[^3P_1]_{6}$        & 1.102     & 0.379     & 0.155     \\ \hline
$[^3P_2]_{6}$        & 0.429     & 0.149     & 0.061     \\ \hline
Total                  & 3.901     &1.345      & 0.550    \\ \hline
\end{tabular}
\label{tab4}
\end{table}
\subsection{ Uncertainty analysis}\label{section:33}
Furthermore, we would like to discuss theoretical uncertainties of the $Z$-boson decay into $\Xi_{QQ'}$, which arise from three main sources: the mass parameters $m_{Q(Q')}$, the transition probability $h_{\bar 3 (6)}$, and the coupling constant. Although the transition probability and the coupling constant might have theoretical uncertainties, they only affect the results as an overall factor. Hence, we will not delve into them here. On the other hand, decay width is greatly influenced by heavy quark masses $m_{Q(Q')}$. Therefore, we varied $m_c = 1.8 \pm 0.3$ and $m_b = 5.1 \pm 0.3$ GeV for doubly heavy baryons with $c$-quark and $b$-quark production to discuss the uncertainties, respectively.  The results are listed in Tables~\ref{tab4}, \ref{tab5}, \ref{tab6} and \ref{tab7}. From the tables, one can observe that
\begin{table}[t]
\centering
\caption{Our result for decay widths (in unit: KeV) of the process $Z$-boson decay into $\Xi_{bc}$ by varying $m_c$ (in unit: GeV).}
\begin{tabular}{|c|c|c|c|}
\hline
State & $m_c$=1.5GeV & $m_c$=1.8GeV & $m_c$=2.1GeV \\ \hline
$[^1P_1]_{\bar3}$  & 2.024     & 0.720     & 0.308     \\ \hline
$[^1P_1]_{6}$        & 1.012     & 0.360     & 0.154     \\ \hline
$[^3P_0]_{\bar3}$  & 1.183     & 0.498     & 0.246     \\ \hline
$[^3P_0]_{6}$        & 0.592     & 0.249     & 0.123     \\ \hline
$[^3P_1]_{\bar3}$  & 2.476     & 0.928     & 0.415     \\ \hline
$[^3P_1]_{6}$        & 1.238     & 0.464     & 0.207     \\ \hline
$[^1P_2]_{\bar3}$  & 2.761     & 0.893     & 0.347     \\ \hline
$[^3P_2]_{6}$        & 1.380     & 0.447     & 0.173     \\ \hline
Total                  & 12.666    & 4.559     & 1.937     \\ \hline
\end{tabular}
\label{tab5}
\end{table}
\begin{table}[t]
\centering
\caption{Our result for decay widths (in unit: KeV) of the process $Z$-boson decay into $\Xi_{bc}$ by varying $m_b$ (in unit: GeV).}
\begin{tabular}{|c|c|c|c|}
\hline
State & $m_b$=4.8GeV & $m_b$=5.1GeV & $m_b$=5.4GeV \\ \hline
$[^1P_1]_{\bar3}$  & 0.736     & 0.720     & 0.707     \\ \hline
$[^1P_1]_{6}$        & 0.368     & 0.360     & 0.353     \\ \hline
$[^3P_0]_{\bar3}$  & 0.536     & 0.498     & 0.466     \\ \hline
$[^3P_0]_{6}$        & 0.268     & 0.249     & 0.233     \\ \hline
$[^3P_1]_{\bar3}$  & 0.946     & 0.928     & 0.896     \\ \hline
$[^3P_1]_{6}$        & 0.482     & 0.464     & 0.448     \\ \hline
$[^3P_2]_{\bar3}$  & 0.881     & 0.893     & 0.905     \\ \hline
$[^3P_2]_{6}$        & 0.441     & 0.447     & 0.452     \\ \hline
Total                  & 4.658     & 4.559     & 4.460     \\ \hline
\end{tabular}
\label{tab6}
\end{table}

\begin{table}[ht]
\centering
\caption{Decay widths (in unit: KeV) for the production of $\Xi_{bb}$ via Z decays by varying $m_b$ (in unit: GeV).}
\begin{tabular}{|c|c|c|c|}
\hline
State & $m_b$=4.8GeV & $m_b$=5.1GeV & $m_b$=5.4GeV \\ \hline
$[^1P_1]_{\bar3}$  & 0.042     & 0.029     & 0.021     \\ \hline
$[^3P_0]_{6}$        & 0.035     & 0.024     & 0.018     \\ \hline
$[^3P_1]_{6}$        & 0.039     & 0.027     & 0.020     \\ \hline
$[^3P_2]_{6}$        & 0.015     & 0.011     & 0.008     \\ \hline
Total                  &0.131     & 0.091      & 0.067       \\ \hline
\end{tabular}
\label{tab7}
\end{table}
\begin{itemize}
\item In comparing Table~\ref{tab4} with Table~\ref{tab7}, it can be see that the influence for $Z\to \Xi_{cc}$ decay widths from $c$-quark mass are greatly larger than $b$-quark mass.
\item From Tables.\ref{tab5} and \ref{tab6}, it shows that the deviation of $m_c$ or $m_b$ has a more greater effect on the decay width for $\Xi_{bc}$ production.
\item The total theoretical prediction by changing $m_c = 1.8 \pm 0.3$ GeV is $\Gamma(\Xi_{cc})= 1.345^{+1.746}_{-0.795}$ KeV and $\Gamma(\Xi_{bc})=4.559^{+8.107}_{-2.622}$ KeV for baryon $\Xi_{cc}$ and $\Xi_{bc}$ production in $P$-wave, respectively.
\item The total theoretical prediction by changing $m_b = 5.1\pm0.3$ GeV is $\Gamma(\Xi_{bc})=4.559^{+0.099}_{-0.099}$ KeV and $\Gamma(\Xi_{bb})=0.091^{+0.040}_{-0.024}$ KeV for baryon $\Xi_{bc}$ and $\Xi_{bb}$ production in $P$-wave, respectively.
\end{itemize}

\section{Summary}\label{section:4}
In this paper, we carry out an investigation of the excited doubly heavy baryons $\Xi_{QQ'}$ production through processes $Z\to\Xi_{QQ'}+\bar Q+\bar Q'$ based on the NRQCD factorization approach. By considering all the contributions from these intermediate $P$-wave diquark states with spin and color quantum number, {\it i.e}., $\langle cc\rangle[^1P_1]_{\bar 3}$, $\langle cc\rangle[^3P_0]_6$, $\langle cc\rangle[^3P_1]_{\bar 3(6)}$, and $\langle cc\rangle[^3P_2]_{\bar 3(6)}$ $\langle bc\rangle[^1P_1]_{\bar 3(b)}$, $\langle bc\rangle[^3P_0]_{\bar 3(6)}$, $\langle bc\rangle[^3P_1]_{\bar 3(6)}$, and $\langle bc\rangle[^3P_2]_{\bar 3(6)}$ $\langle bb\rangle[^1P_1]_{\bar 3}$, $\langle bb\rangle[^3P_0]_6$, $\langle bb\rangle[^3P_1]_6$, and $\langle bb\rangle[^3P_2]_6$, the decay widths for $P$-wave states $\Xi_{cc}$, $\Xi_{bc}$ and $\Xi_{bb}$ are 1.345 KeV, 4.559 KeV and 0.091 KeV, respectively. Additionally, $S$-wave diquark states contributions are also presented for the comprehensive comparison and analysis. Then, we calculated the ratio of $P$-wave diquark states contributions to $S$-wave, which lead to the values $7\%$, $8\%$, $3\%$ for $\Xi_{cc}$, $\Xi_{bc}$, $\Xi_{bb}$ respectively.

The contribution of the $P$-wave diquark states can be taken as the higher-order contribution of $S$-wave states. If all these excited $P$-wave states completely decay into the ground state, we estimate the total decay width
\begin{eqnarray}
&&\Gamma(\Xi_{cc})= 20.220~{\mathrm{KeV}},
\nonumber\\
&&\Gamma(\Xi_{bc})=61.813 ~{\mathrm{KeV}},
\nonumber\\
&&\Gamma(\Xi_{bb})=2.958 ~~{\mathrm{KeV}}.
\end{eqnarray}
After combined $S$-wave and $P$-wave contribution, there will be about $8.104\times 10^3~(10^6)$ $\Xi_{cc}$ events, $24.773\times 10^3~(10^6)$ $\Xi_{bc}$ events and $1.185\times 10^3~(10^6)$  $\Xi_{bb}$ events produced in one operation year at the LHC (CEPC). Meanwhile, the $P$-wave states events will reach to $0.539\times 10^3~(10^6)$, $1.827\times 10^3~(10^6)$, $0.036\times 10^3~(10^6)$ for $\Xi_{cc}$, $\Xi_{bc}$, $\Xi_{bb}$ in one year at the LHC (CEPC) respectively. From our detailed calculations with abundant events and significant branching ratios, the $P$-wave states in doubly heavy baryons make considerable contributions compared with $S$-wave states. Finally, we present the curves for differential decay width with respect to $s_{23}$ and $z$, {\it i.e} $d\Gamma/ds_{23}$ and $d\Gamma/dz$, which demonstrate the properties of $Z$-boson decay into the excited doubly heavy baryons $\Xi_{QQ'}$ processes. It is hoped that our predictions can provide assistance to experimental measurements.

\acknowledgments
We are grateful to the Referees for their valuable comments and suggestions.  This work was supported in part by the National Natural Science Foundation of China under Grant No.12265010, the Project of Guizhou Provincial Department of Education under Grant No.KY[2021]030.

\end{document}